\documentclass[conf]{new-aiaa}
\usepackage[utf8]{inputenc}
\usepackage{graphicx}
\usepackage{amsmath}
\usepackage[version=4]{mhchem}
\usepackage{siunitx}
\usepackage{longtable,tabularx}
\usepackage{hyperref}
\hypersetup{
	colorlinks=true,
	linktoc=all,
	linkcolor=blue,
	citecolor=red
}
\usepackage[mathscr]{eucal}
\usepackage{amssymb,amsfonts}
\usepackage{algorithm,algorithmicx,algpseudocode}
\usepackage{bbm}
\usepackage{graphicx}
\usepackage{subcaption}
\usepackage{mwe}
\usepackage{textcomp}
\usepackage{xcolor}
\usepackage{pdfpages}
\usepackage{soul}
\usepackage{slashbox}

\DeclareMathOperator*{\trace}{\textit{tr}}

\newtheorem{definition}{\bf Definition}[section]
\newtheorem{remark}{\bf Remark}[section]

\newtheorem{program}{\bf Program}[section]

\newcommand{\nnum}{\nonumber}

\definecolor{Blue}{RGB}{88, 105, 225}

\title{Safety Constrained Multi-UAV Time Coordination: A Bi-level Control Framework in GPS Denied Environment
\footnote{This work has been supported by the National Science Foundation (ECCS-1739732 and CMMI-1663460).}}
\author{Wenbin Wan\footnote{Graduate student, Department of Mechanical Science and Engineering, \textit{wenbinw2@illinois.edu}.},
Hunmin Kim\footnote{Postdoctoral Research Associate, Department of Mechanical Science and Engineering, \textit{hunmin@illinois.edu}},
Yikun Cheng\footnote{Graduate student, Department of Mechanical Science and Engineering,
\textit{yikun2@illinois.edu}.},
Naira Hovakimyan\footnote{Professor, Department of Mechanical Science and Engineering, \textit{nhovakim@illinois.edu}, AIAA Fellow.}
}
\affil{University of Illinois at Urbana-Champaign, Urbana, IL 61801}
\author{
Petros G. Voulgaris\footnote{Professor, Department of Mechanical Engineering, \textit{pvoulgaris@unr.edu}.}}
\affil{University of Nevada, Reno, NV 89557}
\author{
Lui Sha\footnote{Professor, Department of Computer Science, \textit{lrs@illinois.edu}.}
}
\affil{University of Illinois at Urbana-Champaign, Urbana, IL 61801}
\begin{document}
\maketitle
\begin{abstract}
Unmanned aerial vehicles (UAVs) suffer from sensor drifts in GPS denied environments, which can cause safety issues.
To avoid intolerable sensor drifts while completing the time-critical coordination task for multi-UAV systems, we propose a safety constrained bi-level control framework.
The first level is  the \textit{time-critical coordination level} that achieves a consensus of coordination states and provides a virtual target which is a function of the coordination state.
The second level  is  the \textit{safety-critical control level} that is designed to follow the virtual target while adapting the attacked UAV(s) at a path re-planning level to support resilient state estimation.
In particular, the \textit{time-critical coordination level} framework generates the desired speed and position profile of the virtual target based on the multi-UAV cooperative mission by the proposed consensus protocol algorithm.
The \textit{safety-critical control level} is able to make each UAV follow its assigned path while detecting the attacks, estimating the state resiliently, and driving the UAV(s) outside the effective range of the spoofing device within the escape time.
The numerical simulations of a three-UAV system demonstrate the effectiveness of the proposed safety constrained bi-level control framework.
\end{abstract} 

\section{Introduction}
In recent years, there has been an increasing interest in multi-UAV systems due to the wide range of applications, including civilian transportation~\cite{8682048}, aerial photography for agriculture~\cite{tik20201st}, searching and rescuing~\cite{10.1145/2750675.2750683}, and other cooperative tasks.
In order to get accurate and reliable state measurements for completing various cooperative tasks safely, the global positioning system (GPS) is the most widely used senor for multi-UAV systems.
However, GPS receivers are potentially vulnerable to various types of attacks, such as blocking, jamming, and spoofing~\cite{warner2003gps}. 
The Vulnerability Assessment Team at Los Alamos National Laboratory has demonstrated that the civilian GPS spoofing attacks can be easily implemented by using GPS simulator~\cite{warner2002simple}.
Furthermore, GPS is more vulnerable when its signal strength is weak.
In particular, due to various applications of multi-UAV systems, the operating environment becomes diverse as well, where GPS signals are weak or even denied due to other structures such as skyscrapers, elevated highways, bridges, and mountains.

\textit{Literature review.} One of the GPS spoofing attack detection techniques is to analyze raw antenna signals or utilize multi-antenna receiver systems.
The GPS spoofing attack can be detected by checking whether the default radiation pattern is changed in~\cite{mcmilin2014single}.
A multi-antenna receiver system was used to detect GPS spoofing attacks by monitoring the angle-of-arrival of the spoofing attempts in~\cite{montgomery2011receiver}.
As an extension of this work, the GPS spoofing mitigation has also been investigated where an array of antennas is utilized to obtain genuine GPS signals by spatial filtering~\cite{magiera2015detection,chen2012study,chen2013validation}.
However, those solutions require modifications of the hardware or the low-level computing modules and assume that an attacker can only use single-antenna spoofing systems.
Furthermore, the attacker can spoof the GPS receivers without being detected if multi-antenna spoofing devices are available~\cite{jansen2017advancing}.

In Cyber-physical system (CPS) security literature, GPS spoofing attacks have been described as a malicious signal injection to the genuine sensor output~\cite{mo2010false}.
Attack detection against malicious signal injection has been widely studied over the last few years.
The attack detection problem has been formulated as an $\ell_0$/$\ell_\infty$ optimization problem, which is NP-hard in~\cite{fawzi2014secure,pajic2014robustness}.
The fundamental limitations of structural detectability, as well as graph-theoretical detectability for linear time-invariant systems, have been studied in~\cite{pasqualetti2013attack}, where distributed attack detection has also been studied.
The attack detection problem has been formulated as an attack-resilient estimation problem of constrained state and unknown input in~\cite{wan2019attack}.
A switching mode resilient detection and estimation framework for GPS spoofing attacks has been studied in~\cite{yoon2019towards}.
Attack detection using multiple GPS signals by checking cross-correlation was introduced in~\cite{psiaki2013gps}.
In~\cite{5718158}, the maximum deviations of the state were identified due to the sensor attacks while remaining stealthy due to the detection.
Resilience to cyber-attacks for multi-agent systems becomes more challenging than for single-agent systems.
There has been much effort in investigating resilient strategies for multi-agent systems in the presence of cyber-attack.
A method of switching the network topologies is utilized to secure consensus tracking performance in the presence of the cyber-attack on communication channels in~\cite{feng2016distributed}.
In ~\cite{ding2016observer}, an event-triggered mechanism and a distributed observer-based controller are designed to ensure the overall consensus of multi-agent systems is achieved.
The coordinated path following design based on an adaptive control method and a synchronization scheme is presented in~\cite{gu2019adaptive}, where coordinated path following goal is achieved.
These architectures can efficiently handle a class of attacks for multi-agent systems, but do not consider fundamental problems indirectly induced by attacks and cannot address the significant problem due to limited sensor availability in the presence of cyber-attacks.

\textit{Contribution.}
The current paper addresses safety problems induced by limited sensor availability due to GPS spoofing attacks while completing the time-critical coordination task for a Multi-UAV system.
We model the sensor drift problem in the presence of GPS spoofing attacks as an increasing variance of state estimation to quantify the sensor drift and introduce \textit{escape time} under which the state estimation error remains within a tolerable error with high confidence.
We propose a safety constrained bi-level control framework for multi-UAV systems that adapts the UAV(s) at a path re-planning level to support resilient state estimation against GPS spoofing attacks.
The proposed framework achieves a consensus of coordination state at the \textit{time-critical coordination level} and is equipped with an escape controller (ESC) that drives the UAV(s) away from the effective range of the spoofing device within the escape time to avoid intolerable sensor drift at \textit{safety-critical control level}.

The remainder of this paper is organized as follows:
In Section~\ref{sec:pre}, we introduce the notation convention, definition of the \textit{escape time} and the dynamic system models for multi-UAV systems.
In the same section, we formulate the problem.
In Section~\ref{sec:framework}, we propose a resilient safety constrained bi-level control framework.
In Section~\ref{sec:simulation}, the numerical simulations of the multi-UAV system for a time-critical mission under the GPS spoofing attack is presented.
Section~\ref{sec:conclusion} draws the conclusion.

\section{Preliminaries}\label{sec:pre}

\subsection{Notation}
We use the subscript $k$ of $x_k$ to denote the time index;
${\mathbb R}^n_+$ denotes the set of positive elements in the $n$-dimensional Euclidean space;
${\mathbb R}^{n \times m}$ denotes the set of all $n \times m$ real matrices;
$A^\top$, $\trace{(A)}$ and $A^{-1}$ denote the transpose, trace and inverse of matrix $A$, respectively;
$I$ denotes the identity matrix with an appropriate dimension;
$\|\cdot\|$ denotes the standard Euclidean norm for a vector or an induced matrix norm;
$\times$ is used to denote Cartesian product;
${\mathbb E}[\,\cdot\,]$ denotes the expectation operator.
For a matrix $S$, $S > 0$ and $S \geq 0$ indicate that $S$ is positive definite and positive semi-definite, respectively.

\subsection{Escape time}

In the presence of the GPS spoofing attack, the state estimation algorithm relies on the relative measurement sensors because the GPS signals do not contain legitimate information.
In this case, the variance of the state estimation errors is strictly increasing and unbounded in time (Theorem 4.2 in~\cite{yoon2019towards}).
Regarding the sensor drift problem, we utilize a new resilience measure, escape time, which is defined as follows:
\begin{definition}\cite{yoon2019towards}
The escape time $k^{esc} \geq 0$ is the time difference between the attack time $k^a$ and the first time instance when the estimation error $\|x_k-\hat{x}_k\|$ is not within the tolerable error distance $\zeta \in {\mathbb R}^n_+$ with the significance $\alpha$, i.e.
\begin{align*}
    &k^{esc}=\arg \min_{k \geq k^a} k - k^a\nnum\\
    &\text{s.t.  } \zeta^\top P_{k}^{-1}\zeta < \chi_{df}^2(\alpha), \label{escapetime}
\end{align*}
where $P_{k}$ is the error covariance of $x_k-\hat{x}_k$, and $\chi^2$ is the chi-squared test value with degree of freedom $df$.
\label{def1}
\end{definition}
The escape time provides a new safety criterion for optimal control with increasing uncertainties.
It is worth to notice that the escape time $k^{esc}$ can be calculated by Algorithm 1 in~\cite{yoon2019towards}.

\subsection{System model}

In what follows, we describe the multi-UAV system in detail.

\subsubsection{Agent model}
Consider the discrete-time dynamic system model of the \textit{single agent}:
\begin{subequations} \label{eq1 sys}
\begin{align} 
    x_{k+1} &= A x_{k} + B u_{k} + w_{k} \\
    y_k^G &= C^G x_k + d_k + v_k^G \\
    y_k^I &= C^I (x_k - x_{k-1}) + v_k^I
            \end{align}
\end{subequations}
where $x_k \in \mathbb{R}^n$ is the state, $u_k\in \mathbb{R}^m$ is the control input and $A$, $B$, $C^G$ and $C^I$ are the system matrix, input matrix, and output matrix with proper sizes.
The sensor measurement $y_k^G \in {\mathbb R}^{m_G}$ is the GPS measurement which may be corrupted by unknown GPS spoofing signal $d_k \in {\mathbb R}^{m_G}$.
We assume that the attacker can inject any signal $d_k$ into $y_k^G$.
The sensor measurement $y_k^I \in {\mathbb R}^{m_I}$ is the inertial measurement unit (IMU) measurement which returns a noisy measurement of the state difference.
The output $y_k^I$ can represent any  {\em relative} sensor measurement, such as velocity measurement by a camera. In this paper, we use IMU for the illustration.

The noise signals $w_k$, $v_k^G$ and $v_k^I$ are assumed to be independent and identically distributed (i.i.d.) Gaussian random variables with zero means and covariances
${\mathbb E[w_k w_k^\top]=\Sigma}_w \geq 0$,
${\mathbb E[v_k^G (v_k^G)^\top]=\Sigma}_G >0$,
and ${\mathbb E[v_k^I (v_k^I)^\top]=\Sigma}_I>0$, respectively.

\subsubsection{Multi-agent network} \label{multi_agent_sys}
Let $x_{i,k} \in \mathbb{R}^n$, $i = 1, \cdots, N$ be the state of the $i^{th}$ agent associated with dynamic system model~\eqref{eq1 sys} where $N$ is the total number of the agents.
Graph theory can provide the natural abstractions for how information is shared between agents in a network~\cite{mesbahi2010graph}.
An undirected graph $\mathcal{G} = (V,E)$ consists of a set of nodes $V = \{1, 2, \cdots,N\}$, which corresponds to the different agents, and a set of edges $E \subset V\times V$, which relates to a set of unordered pairs of agents.
In particular, $(i,j), (j,i)\in E$ if and only if there exists a communication channel between agents $i$ and $j$.
The neighborhood $\mathcal{N}(i) \subseteq V$ of the agent $i$ will be understood as the set $\{j \in V \mid (i,j) \in E\}$.

\subsubsection{Path following consensus} \label{path following consensus}

\begin{figure}[thpb]
\centering
 \includegraphics[width=0.6\textwidth]{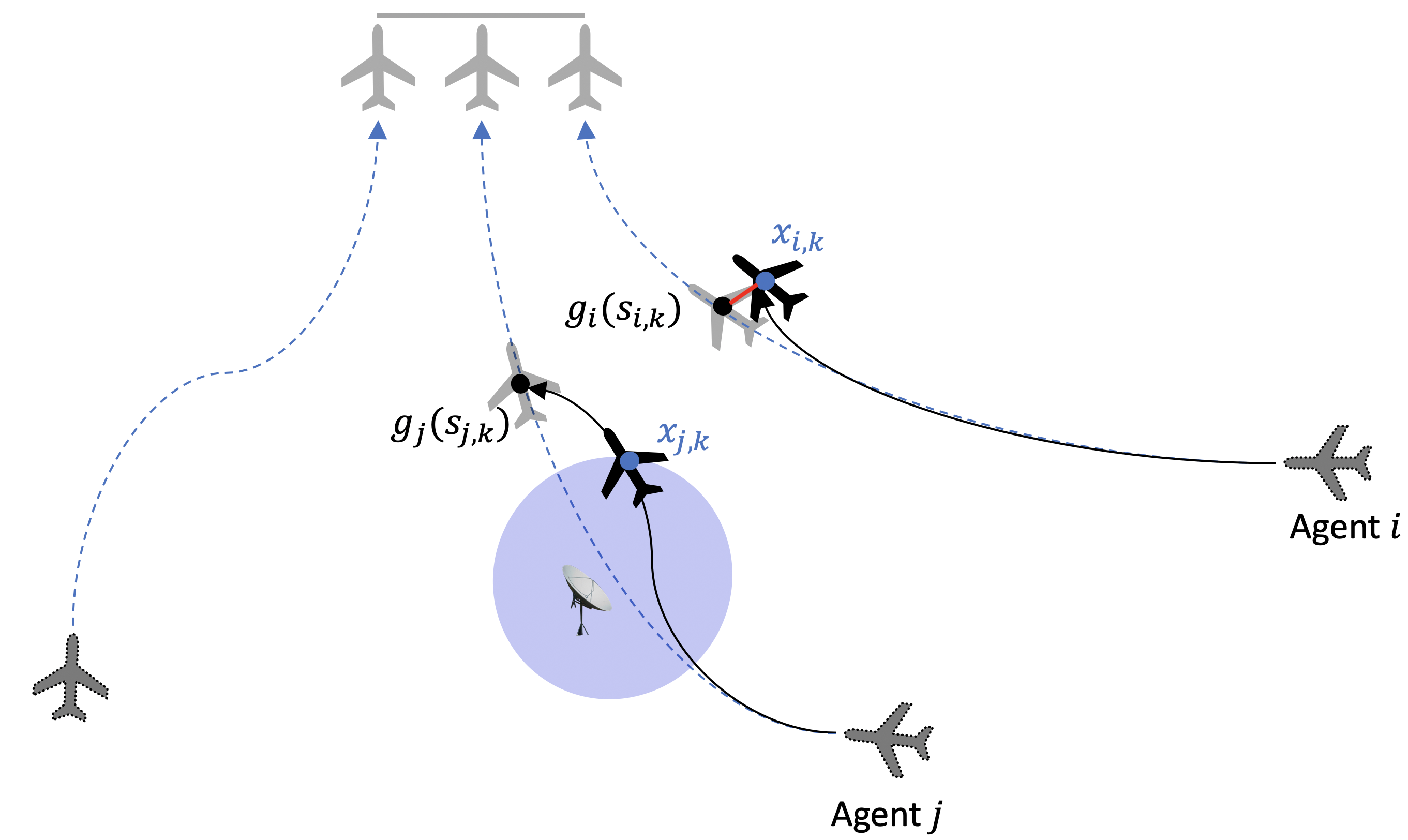}
\caption{Illustration of the path following consensus. The goal of the multi-agent system is for all agents to reach  the desired goal state simultaneously. For the agent $i$ at time $k$, the virtual target/predetermined desired state is $g_i(s_{i,k})$ and the true state is $x_{i,k}$.
The error between the virtual target $g_i(s_{i,k})$ and the true state $x_{i,k}$ (marked in red) is to be minimized.
The attacker is on the path of the agent $j$ and the effective spoofing area is displayed as the light blue circle.
When the attack is detected, the agent $j$ will be re-planning the trajectory so that the state estimation errors remain in the tolerable region, while the other agents will adjust their coordination states accordingly to achieve time-coordination.}
\label{fig:demo}
\end{figure}

Each agent $i \in V$ has a desired trajectory $g_i: s_{i,k} \rightarrow {\mathbb R}^{n_s}$ that is parameterized by coordination state variable $s_{i,k} \in [0,1]$ as shown in Fig.~\ref{fig:demo}. Dimension $n_s$ is usually $2$ ($2-$D mission) or $3$ ($3-$D mission).
At time $k$, $g_i(s_{i,k})$ is the virtual target that the agent $i$ follows at that time, i.e., agent $i$ pursues to minimize the error $\|g_i(s_{i,k}) - x_{i,k}\|$ which is marked in red in Fig.~\ref{fig:demo}.
The state $s_{i,k}$ can be seen as a normalized length of trajectory.
The agents also desire to achieve the consensus of the coordination state variable
\begin{align*}
    s_{i,k} - s_{j,k} \stackrel{k \rightarrow \infty}{\longrightarrow} 0 \quad \forall i,j \in V,
\end{align*}
so that the virtual targets of the agents arrive at the destination at the same time.



The agent $i$ knows coordination state $s_{i,k}$ as well as the coordination states $s_{j,k}$ for neighboring agents $j \in \mathcal{N}(i)$.

\subsection{Problem Statement}\label{sec:pro} 
Given a multi-agent network described in Section~\ref{multi_agent_sys} consisting of number of $N$ agents described in~\eqref{eq1 sys}, the agent $i$, where $i = 1, \cdots, N$, aims to follow its desired trajectory $g_i(\cdot)$ with a reference rate $\rho$, i.e., 
\begin{subequations} \label{eq progressing rate}
\begin{align} 
    g_i(s_{i,k}) &-x_{i,k} \stackrel{k \rightarrow \infty}{\longrightarrow} 0\\
    s_{i,k+1} &- s_{i,k} \stackrel{k \rightarrow \infty}{\longrightarrow}  \rho,
\end{align}
\end{subequations}
and to achieve time coordination, i.e.,
\begin{align} \label{eq consensus prob}
    s_{i,k} - s_{j,k} \stackrel{k \rightarrow \infty}{\longrightarrow} 0
\end{align}
for all $i,j \in V$, and for all $k \geq 0$.
Meanwhile, each agent aims to detect the GPS spoofing attack;
obtain the attack-resilient state estimation when considering the limited sensor availability;
complete the path following mission securely.

\section{Safety Constrained Bi-level Control Framework}\label{sec:framework}

\begin{figure}[thpb]
\centering
 \includegraphics[width=0.6\textwidth]{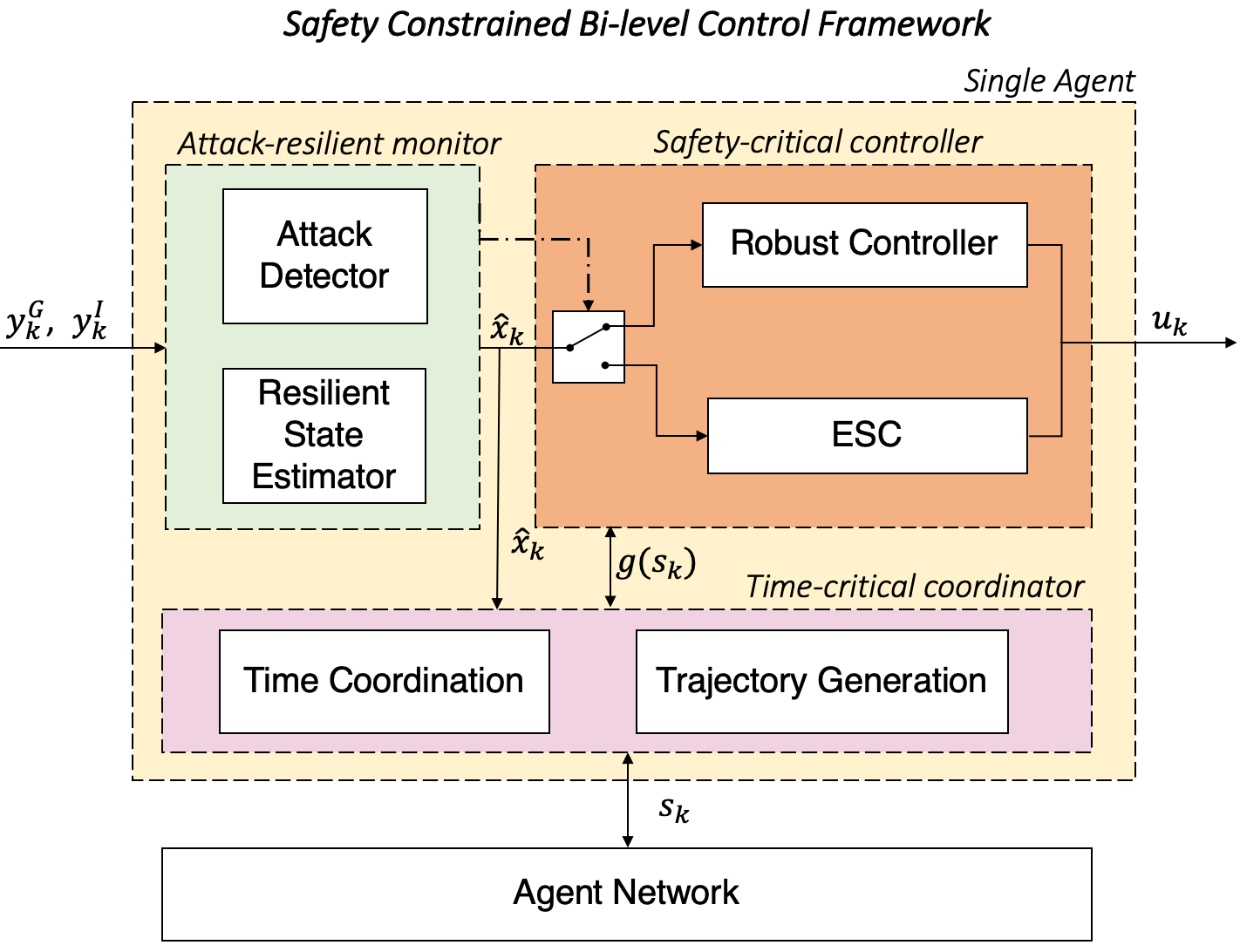}
\caption{A safety constrained bi-level control framework consisting of an attack-resilient monitor, a safety-critical controller and a time critical coordinator.}
\label{fig:framework}
\end{figure}

To address the problem described in Section~\ref{sec:pro} for a multi-agent system, we propose a safety constrained bi-level control framework shown in Fig.~\ref{fig:framework}.
The first level, \textit{time-critical coordination level}, is designed to achieve time coordination with the assigned trajectories.
The second level is at \textit{safety-critical control level} that supports resilient estimation and path following control.
The safety constrained bi-level control framework consists of a time-critical coordinator at the \textit{time-critical coordination level}; an attack detector, a resilient state estimator, a robust controller, and an escape controller (ESC) at the \textit{safety-critical control level}.

The following explains each module in the proposed framework as shown in Fig.~\ref{fig:framework}.


\textbf{Time-critical coordinator.}
The time coordination algorithm guarantees that each agent reaches an agreement on some distributed variables of interest; i.e., the coordination state variables.
With the coordination state variables, the assigned trajectories generate the virtual targets for UAVs to follow.

\textbf{Safety-critical controller.}
The robust controller is a complex controller that operates the UAV to follow the virtual targets in the presence of noise, but without the presence of attacks.
The robust controller can be implemented as any effective control technique such as optimal control, model predictive control, PID, etc.
The escape controller (ESC) is a model predictive controller (MPC)-based structure that adapts the UAV at a path re-planning level for safe operation.
ESC drives the UAV out of the effective range of the spoofing device \textit{within the escape time}.

\textbf{Attack-resilient monitor.}
The resilient state estimator is developed based on the Kalman-filter like state estimator.
The attack detector is designed by the $\chi^2$-based anomaly detection algorithm.
Based on the previous estimation from the resilient state estimator, the Boolean output (the dot-dashed line in Fig.~\ref{fig:framework}) of the attack detector determines $i)$ whether the GPS measurement should be used for the state estimation and $ii)$ the switching rule between the two controllers: the robust controller and the escape controller (ESC).

In what follows, each subsection describes the details of the corresponding component.

\subsection{Time coordination (Consensus Protocol)}

Consider the coordinate state of the consensus network model
\begin{align} \label{eq consensus network}
s_{i,k+1} = s_{i,k}  + z_{i,k},
\end{align}
where $z_{i,k} \geq 0 $ is the control input for the coordination state of the agent $i$ at time index $k$.
To solve the path following consensus problem in~\eqref{eq progressing rate} and~\eqref{eq consensus prob}, we propose the design of the control input $z_{i,k}$.
The control input $z_{i,k}$ is designed by
\begin{align}\label{eq rb control}
    z_{i,k} = \max\left\{-k_e\|g_i(s_{i,k})-x_{i,k}\| -k_s \sum_{j \in \mathcal{N}(i)}(s_{i,k}-s_{j,k}) + \rho + \mathbbm{1} _{\text{attacked}}\hat{z}_{i,k}, 0 \right\},
\end{align}
where $k_e > 0$ and $k_s>0$ are coordination control gains, and the reference rate $\rho$ is the desired rate of progress that is a constant.
The first term $-k_e\|g_i(s_{i,k})-x_{i,k}\|$ indicates that the agent reduces the coordination speed when there is a tracking error.
The second term $-k_s \sum_{j \in \mathcal{N}(i)}(s_{i,k}-s_{j,k})$ is the consensus term which reduces errors between the local coordination state with those of the neighbors.
The third term $\rho$ is the desired rate if there is no tracking error and no coordination error.
The last term $\hat{z}_{i,k} = k_e\|g_i(s_{i,k})-x_{i,k}\|$ drives the virtual target away from the spoofing device even when the UAV detours the planned trajectory.
Function $\mathbbm{1} _{i,\text{attacked}}$ is an indicator function and  $\mathbbm{1} _{i,\text{attacked}} = 1 $ if an attack is detected, otherwise $\mathbbm{1} _{i,\text{attacked}} = 0 $.
Moreover, if $-k_e\|g_i(s_{i,k})-x_{i,k}\| -k_s \sum_{j \in \mathcal{N}(i)}(s_{i,k}-s_{j,k}) + \rho + \mathbbm{1} _{i,\text{attacked}}\hat{z}_{i,k}$ is less than zero, then the virtual target chooses to stay at current state rather than go backwards.

\subsection{Resilient State Estimator}\label{sec:estimation}
The defender implements an estimator and $\chi^2$ detector to estimate the state and detect the GPS spoofing attack.
The following Kalman-filter like state estimator is used to estimate the current state:
\begin{align}
\hat{x}_{k} &= A \hat{x}_{k-1}+ B u_{k-1}
+ K_k^G (y_k^G - C^G( A\hat{x}_{k-1} + B u_{k-1}))+ K_k^I (y_k^I  -C^I( A \hat{x}_{k-1} + B u_{k-1}-\hat{x}_{k-1})) \label{eq x estim}\\
P_k&= (A-K_kCA+K_kDC)P_{k-1}
 (A-K_kCA+K_kDC)^\top +(I-K_kC)\Sigma_w (I-K_kC)^\top+K_k \Sigma_y K_k^\top, \label{eq P_update}
\end{align}
where $\hat{x}_k$ is the state estimate and $P_k$ is the state estimation error covariance at time $k$.
We define
\begin{align*}
    K_k :=\left[ \begin{array}{cc} K_k^G& K_k^I \end{array} \right], \quad C:= \left[ \begin{array}{c}  C^G\\ C^I\\ \end{array} \right], \quad  \Sigma_y := \left[ \begin{array}{cc} \Sigma_G& 0\\ 0&\Sigma_I\\ \end{array} \right],\quad \text{and} \quad D := \left[ \begin{array}{cc} 0&0\\ 0&I\\ \end{array} \right].
\end{align*}
The optimal gain $K_k$, given by
\begin{align}\label{eq:K_law}
&K_k=(AP_{k-1}(CA-DC)^\top+\Sigma_w C^\top) \left((CA-DC)P_{k-1}(CA-DC)^\top\right.+\left.C\Sigma_wC^\top+\Sigma_y\right)^{-1},\nonumber
\end{align}
is the solution of the optimization problem $\min_{K_k} \trace{(P_k)}$.

In~\cite{yoon2019towards}, it has been shown that the covariance in~\eqref{eq P_update} is bounded when the GPS signal is available. If the GPS is denied, and only the relative sensor $y_k^I$ is available, the covariance is strictly increasing and is unbounded in time. That is, the sensor drift problem can be formulated as  instability of the covariance matrix.

\subsection{Attack Detector}
We conduct the $\chi^2$ test to detect the GPS spoofing attacks:
\begin{equation}
     H_0: d_k=0; \quad H_1: d_k \neq 0, \label{e000.1}
\end{equation}
using CUSUM (CUmulative SUM) algorithm, which is widely used in attack detection research~\cite{page1954continuous,barnard1959control,lai1995sequential}.

Since $d_k = y_k^G - C^G x_k - v_k^G$, given the previous state estimate $\hat{x}_{k-1}$, we estimate the attack vector by comparing the sensor output and the output prediction:
\begin{align}
    \hat{d}_k     &=y_k^G - C^G ( A\hat{x}_{k-1} + B u_{k-1}). \label{est d_k}
\end{align}
Note that the current estimate $\hat{x}_k$ should not be used for the prediction, because it is correlated with the current output; i.e., ${\mathbb E}[\hat{x}_k (y_k^G)^\top] \neq 0$.
Due to the Gaussian noises $w_k$ and $v_k$ injected to the linear system in~\eqref{eq1 sys}, the states follow Gaussian distribution since any finite linear combination of Gaussian distributions is also Gaussian.
Similarly, $\hat{d}_k$ is Gaussian as well, and thus the use of $\chi^2$ test~\eqref{e000.1} is justified.
In particular, the $\chi^2$ test compares the normalized attack vector estimate $\hat{d}_k^\top (P_{k}^d)^{-1}\hat{d}_k$ with $\chi^2_{df}(\alpha)$:
\begin{equation}
\begin{aligned}
&\text { Accept $H_0$, if } \hat{d}_k^\top (P_{k}^d)^{-1}\hat{d}_k \leq \chi^2_{df}(\alpha)\\
&\text { Accept $H_1$, if } \hat{d}_k^\top (P_{k}^d)^{-1}\hat{d}_k > \chi^2_{df}(\alpha),
\end{aligned}\label{e003}    
\end{equation}
where $P_{k}^d := {\mathbb E}[(d_k-\hat{d}_k)(d_k-\hat{d}_k)^\top]=C^G(A P_{k-1}A^\top+\Sigma_w)(C^G)^\top+\Sigma_G$, and $\chi_{df}^2(\alpha)$ is the threshold found in the Chi-square table. In $\chi_{df}^2(\alpha)$, $df$ denotes the degree of freedom, and $\alpha$ denotes the statistical significance level.

To reduce false positive/negative due to noise, we use the test~\eqref{e003} in a cumulative form.
The proposed $\chi^2$ CUSUM detector is characterized by the detector state $S_k\in\mathbb{R}_{+}$:
\begin{align}
    S_{k}=\delta S_{k-1}+(\hat{d}_k)^\top (P_{k}^d)^{-1}\hat{d}_k, \quad S_0=0,
    \label{e003.1}
\end{align}
where $0<\delta<1$ is the pre-determined forgetting factor. 
At each time $k$, the CUSUM detector~\eqref{e003.1} is used to update the detector state $S_k$ and detect the attack.

The attack detector will $i)$ update the estimated state $\hat{x}_k$ and the error covariance $P_k$ in~\eqref{eq P_update} with $K_k^G=0$ and $ii)$ switch the controller to ESC,
if
\begin{align}
     S_k>\sum_{i=0}^{\infty}\delta^i\chi^2_{df}(\alpha)=\frac{\chi^2_{df}(\alpha)}{1-\delta}.\label{e003.2}
\end{align}
If $S_k<\frac{\chi^2_{df}(\alpha)}{1-\delta}$, then it returns to the robust control mode. 

\begin{figure}[htpb]
\centering
 \includegraphics[width=0.5\textwidth]{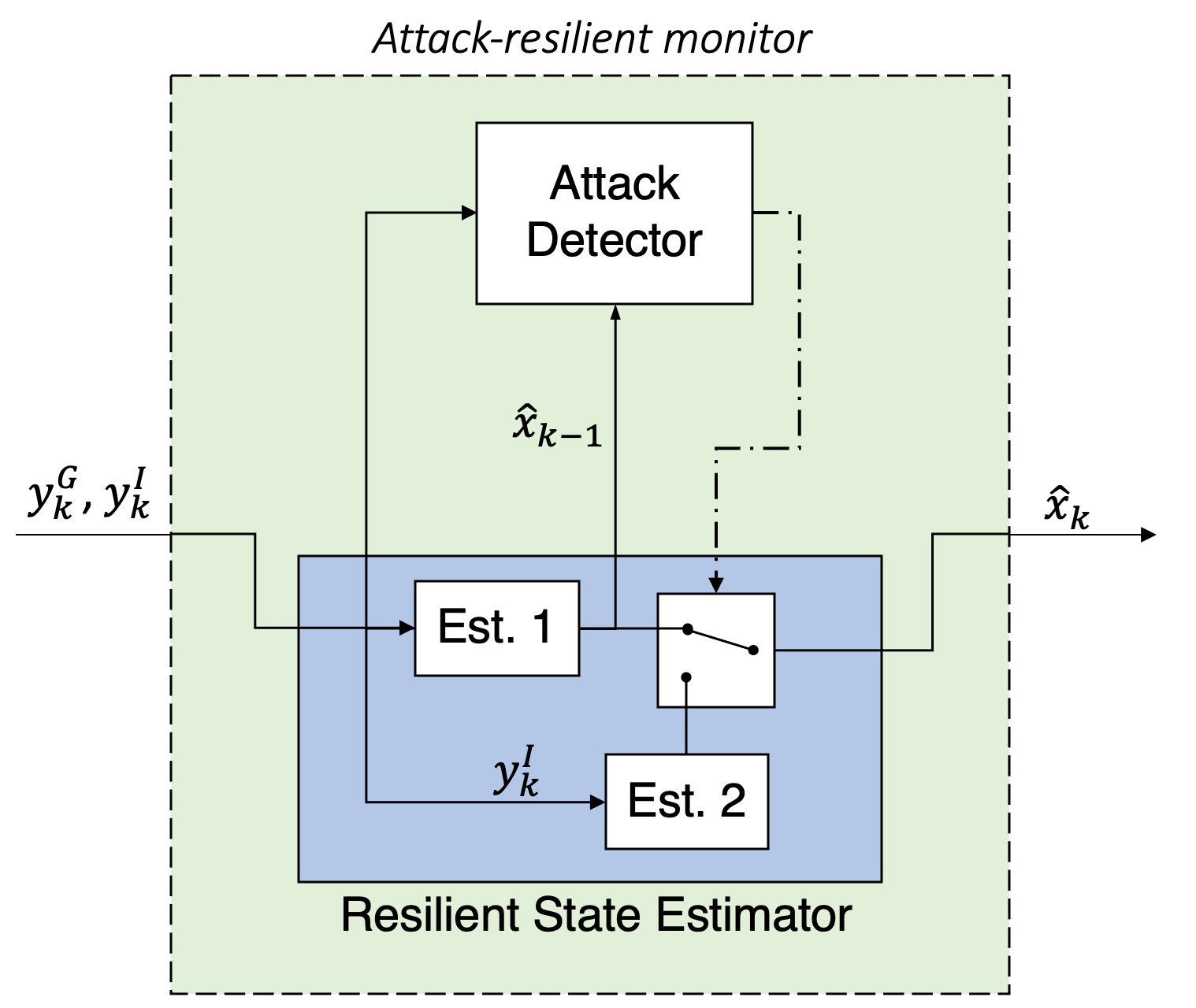}
\caption{ Resilient state estimator. GPS and IMU measurements are used in the Estimator 1 (Est. 1). Estimator 2 (Est. 2) only uses the IMU measurement. Est. 1 is  used to estimate the state by~\eqref{eq x estim} for the detection as in~\eqref{est d_k}. When GPS is free of attacks, Est. 1 is also used to estimate the state for the control since the GPS measurement is trustful. In the presence of the GPS attack, Est. 2 is used for the control.}
\label{fig:est details}
\end{figure}

\begin{remark}
As shown in Fig.~\ref{fig:est details}, the resilient state estimation uses the GPS measurement and the IMU measurement to estimate the state by~\eqref{eq x estim} for the detection purpose as in~\eqref{est d_k}.
When the GPS attack is detected, only the IMU measurement is used to estimate the state for the control purpose as in~\eqref{eq x estim} and~\eqref{eq P_update} with $K^G_k =0$.
\end{remark}

\subsection{Escape Controller (ESC)}

In the presence of the GPS spoofing attack, the variance $P_k$ in~\eqref{eq P_update} of the state estimation errors is strictly increasing and unbounded in time (Thm. 4.2~\cite{yoon2019towards}), and the escape time provides a new criterion for optimal trajectory regeneration with increasing uncertainties.
The goal of ESC is to drive the UAV outside of the effective range of the spoofing device within the escape time so that the state estimation error remains within the tolerable region with a predetermined probability.
In particular, the \textit{safety constraint} can be formulated as
\begin{align}  \label{eq2 mpc01 c}
    d(x^a_{k^a+k^{esc}}, x_{k^a+k^{esc}})-{r_{\textit{effect}}} > 0,
\end{align}
where $x^a_k$ denotes the location of the attacker at time $k$, $k^a$ and $k^{esc}$ are the attack time and the escape time,
and the function $d(a,b)$ measures the distance between $a$ and $b$.
The value $r_{\textit{effect}}$ is the upper bound of the effective range of the spoofing device.
The \textit{safety constraint}~\eqref{eq2 mpc01 c} implies that ESC should drive the UAV outside of the effective range of the spoofing device within the escape time.

\begin{remark}
We assume that the upper bound of the  effective range $r_{\textit{effect}}$ and the location of the attacker $x^a_k$ are known.
Due to hardware constraints, the output power/nominal strength of the spoofing device is bounded where the output power determines the effective range of the spoofing device.
The distance between the attacker and UAV can be obtained by monitoring the injected GPS signal strength using Friis transmission equation ~\cite{1697062}.
The location of the attacker can be estimated similar to  locating the epicenter of an earthquake, which can be done with at least three measurements from different seismic stations by measuring a series of GPS signal strengths from different locations of the UAV.
\end{remark}

There are two key challenges for considering the \textit{safety constraint}~\eqref{eq2 mpc01 c}.
First, the states and the attacker location are unknown, and their estimates $\hat{x}_{i}$ and $\hat{x}^a_{i}$ are subject to stochastic noise.
Moreover, we cannot guarantee that the \textit{safety constraint}~\eqref{eq2 mpc01 c} is always feasible.
Addressing the above two challenges, we replace the \textit{safety constraint}~\eqref{eq2 mpc01 c} by the repulsive potential function~\cite{ge2000new} as a high penalty in the cost function which is active only after the escape time $k^a + k^{esc}$.
The repulsive potential function $U_{rep}(D)$ is defined as the following:
\begin{align*}
    U_{rep}(D) :=\left\{\begin{array}{ll}{\frac{1}{2}\beta\left(\frac{1}{D}-\frac{1}{r_{\textit{effect}}}\right)^{2}} & {\text { if } D \leq r_{\textit{effect}}} \\
    {0} & {\text { if }  D>r_{\textit{effect}}}\end{array}\right.,
\end{align*}
which can be constructed based on the distance between the location of the attacker and the location of UAV, $D := d(x_{k^a+k^{esc}}^a, \hat{x}_{k^a+k^{esc}})$.
The scaling parameter $\beta$ is a large constant, which represents a penalty when the constraint has not been fulfilled.
Utilizing the soft constraint, we reformulate the MPC problem as  follows:
\begin{program}\label{mpc04}
\begin{align}
    \min_{u} \ &\sum_{i=k^a}^{k^a+N}\hat{\tilde{x}}_{i+1}^\top Q_i \hat{\tilde{x}}_{i+1}+u_i^\top R_i u_i + \sum_{i=k^a+k^{esc}}^{k^a+N}U_{rep}(D_i) \nonumber\\
    \text{s.t.  } &\hat{x}_{i+1} = A \hat{x}_i + B u_i \nonumber\\
    &h(\hat{x}_i,u_i) \leq 0 \label{pfinMPC}\\
    &\text{for \ } i = k^a, k^a+1, \cdots, k^a+N, \nonumber
\end{align}
\end{program}
where $N \geq k^{esc}$ is the prediction horizon,
$\hat{\tilde{x}}_i$ is defined as the difference between the state estimation and the goal state at time index $i$, i.e., $\hat{\tilde{x}}_i := \hat{x}_i - x_i^{goal}$,
$Q_i$ and $R_i$ are symmetric positive definite weight matrices,
and $\hat{x}^a_i$ is the estimate of the attacker location.
Value $r_{\textit{effect}}$ is the upper bound of the effective range of the spoofing device.
Inequality~\eqref{pfinMPC} is any nonlinear constraint on the state estimation $\hat{x}_i$ (e.g., velocity) and the control input $u_i$ (e.g., acceleration).

\begin{remark}
Each agent obtains the attacker's information and switches to ESC when it is under attack.
In the cases that a large portion of the planned trajectory is inside the effective range of the spoofing device, following the virtual target may cause the agent to re-enter the effective range when the agent switches back to the robust controller.
Once the agent obtains the attacker information, it will share with the robust controller to avoid re-entering.
\end{remark}

\begin{remark} 
Comparing to the use of the repulsive potential function $U_{rep}$ in the collision avoidance literature~\cite{olfati2006flocking,choset2005principles,wolf2008artificial}, the proposed application of the repulsive potential function in Program~\ref{mpc04} has two differences.
First of all, the repulsive potential function is known before the collision happens in collision avoidance literature, while we can only get the repulsive potential function $U_{rep}$ after the collision happens, i.e., only after the UAV has entered the effective range of the spoofing device.
Second, the repulsive potential function $U_{rep}$ is only counted in the cost function in Program~\ref{mpc04} after the escape time.
\end{remark} 
\section{Simulation}\label{sec:simulation}
The scenario in Fig.~\ref{fig:demo} is used to demonstrate the efficacy of the proposed framework.
In the simulation, three UAVs are moving to the desired goal positions simultaneously from different initial locations by using feedback control\footnote{We implemented a proportional-derivative (PD) like tracking controller, which is widely used for double integrator systems.} based on the state estimate from~\eqref{eq x estim}.
When one of the UAVs is in the effective range of the spoofing device, its state estimate will be no longer trustful.
After the GPS measurement is turned off, the only available relative state measurement causes the sensor drift problem~\cite{yoon2019towards}.
The UAV will switch the controller from the robust controller to ESC when the attack is detected, using ESC to escape away from the attacker within the escape time, while all  UAVs will adjust their coordination states if necessary to achieve time-coordination.
The online computation of ESC is described in Program~\ref{mpc04} done using \texttt{Julia}, and ESC is implemented by using \texttt{JuMP}~\cite{DunningHuchetteLubin2017} package  with \texttt{Ipopt} solver.

\subsection{Single UAV Model}
We use a double integrator UAV dynamics under the GPS spoofing attack as in~\cite{kerns2014unmanned}.
The discrete time state vector $x_k$ considers planar position and velocity at time step $k$, i.e.
\begin{equation*}
    x_k = [r_k^x, r_k^y, v_k^x, v_k^y]^\top,
\end{equation*}
where $r_k^x$, $r_k^y$ denote $x$, $y$ position coordinates, and $v_k^x, v_k^y$ denote velocity coordinates.
We consider the acceleration of UAV as the control input $u_k = [u_k^x, u_k^y]^\top$.
We assume that the state constraint and control input constraint are given as
\begin{align*}
    \sqrt{(v_k^x)^2 + (v_k^y)^2}  \leq 5, \quad \sqrt{(u_k^x)^2 + (u_k^y)^2}  \leq 2.
\end{align*}
With sampling time at $0.1$ seconds, the double integrator model is discretized into the following matrices:
\begin{equation*}
    A = 
    \begin{bmatrix} 
    1 & 0 & 0.1 & 0    \\
    0 & 1 & 0    & 0.1 \\
    0 & 0 & 1    & 0 \\
    0 & 0 & 0    & 1
    \end{bmatrix},
    \quad
    B = 
    \begin{bmatrix} 
    0    & 0 \\
    0    & 0 \\
    0.1 & 0 \\
    0    & 0.1 
    \end{bmatrix},
\end{equation*}
and the outputs $y^G_k$ and $y^I_k$ are the position measurements from GPS and the velocity measurements from IMU, with the output matrices:
\begin{align*}
    C^G = 
    \begin{bmatrix} 
    1 & 0 & 0 & 0    \\
    0 & 1 & 0 & 0
    \end{bmatrix}, \quad
    C^I = 
    \begin{bmatrix} 
    0 & 0 & 1 & 0 \\
    0 & 0 & 0 & 1 
    \end{bmatrix}.
\end{align*}
The covariance matrices of the sensing and disturbance noises are chosen as
$\Sigma_w = 0.1I$, $\Sigma_G = I$ and  $\Sigma_I = 0.01I$.

\subsection{Trajectory generation and time coordination for multi-UAV systems}
The nominal trajectories of a three-UAV system $g_i(s_{i,k})$, where $i \in \{1, 2,3\}$, are generated by the cubic B\'ezier curves \cite{Beziercurve}
\begin{equation}\label{eq:Bezier curve}
    g_i(s_{i,k}) \triangleq (1-s_{i,k})^{3}\mathbf{P}^{(0)}_{i}+3(1-s_{i,k})^{2} s_{i,k} \mathbf{P}^{(1)}_{i} + 3(1-s_{i,k})s_{i,k}^{2}\mathbf{P}^{(2)}_{i}+s_{i,k}^{3}\mathbf{P}^{(3)}_{i},
\end{equation}
where $s_{i,k} \in [0, 1]$ is the coordination state and $\mathbf{P}^{(j)}_{i}$, where $j \in \{0, 1, 2, 3\}$, are the control points for the agent $i$.
The control points we used are listed in Table~\ref{table:control points}.

\begin{table}[ht]
\centering
\begin{tabular}{|c||*{4}{c|}}\hline
\backslashbox{$i$}{$(j)$}
&\makebox[3em]{(0)}&\makebox[3em]{(1)}&\makebox[3em]{(2)}
&\makebox[3em]{(3)}\\\hline\hline
$1$ & $[0\ ,\ \ 0]^\top$ &$[100,\ 100]^\top$&$[10,\ 300]^\top$&$[190,\ 400]^\top$\\\hline
$2$ & $[200,\ 0]^\top$ &$[100,\ 100]^\top$&$[250,\ 200]^\top$&$[200,\ 400]^\top$\\\hline
$3$ & $[400,\ 0]^\top$ &$[450,\ 150]^\top$&$[300,\ 300]^\top$&$[210,\ 400]^\top$\\\hline
\end{tabular}
\caption{B\'ezier curve control points $\mathbf{P}^{(j)}_{i}$}
\label{table:control points}
\end{table}

\begin{figure}[ht]
     \centering
     \begin{subfigure}[b]{0.4\textwidth}
         \centering
         \includegraphics[width=\textwidth]{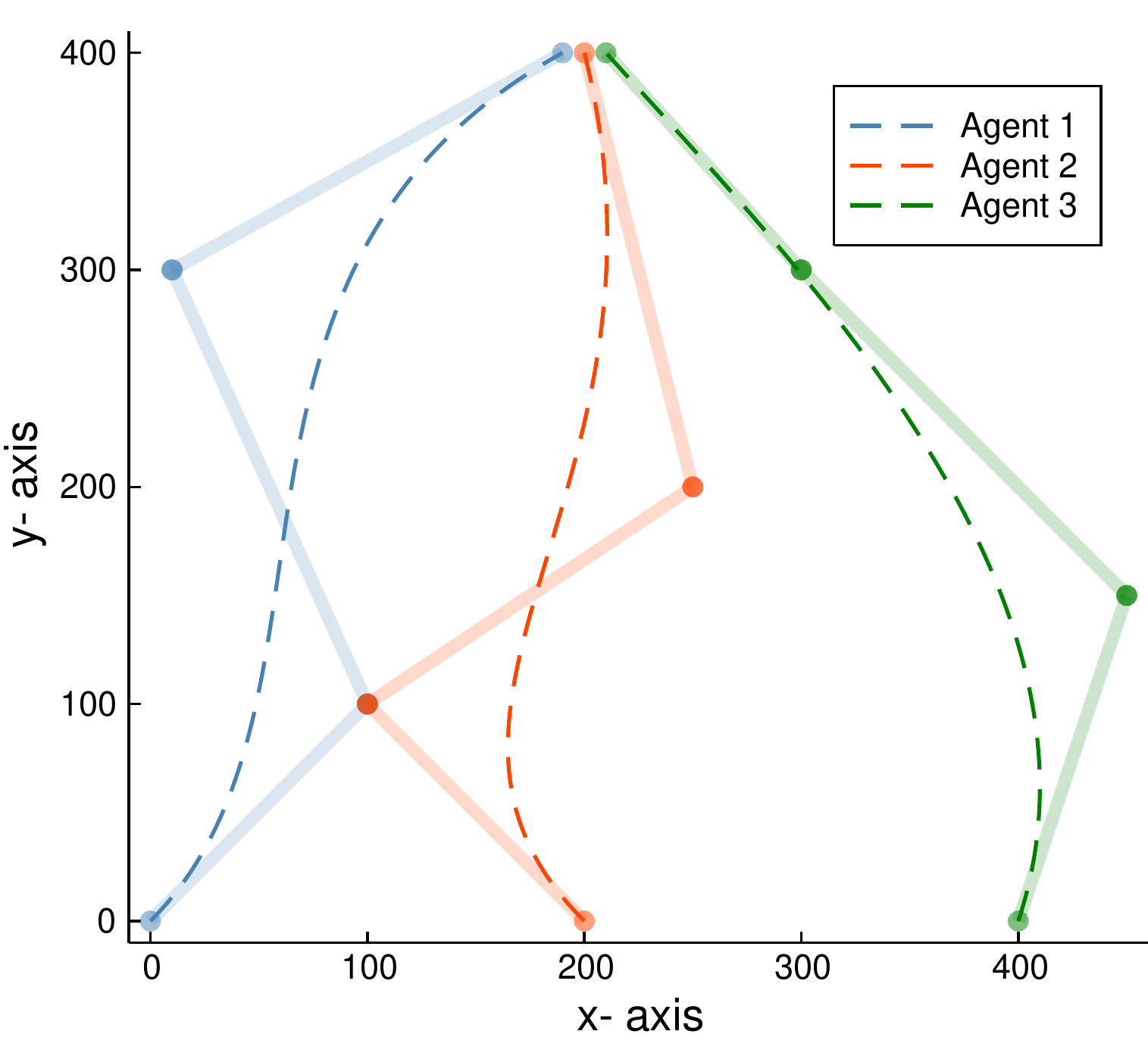}
         \caption{Trajectories of the three agents in dashed lines generated by B\'ezier curves~\eqref{eq:Bezier curve}  using the control points summarized in Table~\ref{table:control points}. }
         \label{fig:bezier traj generation}
     \end{subfigure}
     \hfill
     \begin{subfigure}[b]{0.4\textwidth}
         \centering
         \includegraphics[width=\textwidth]{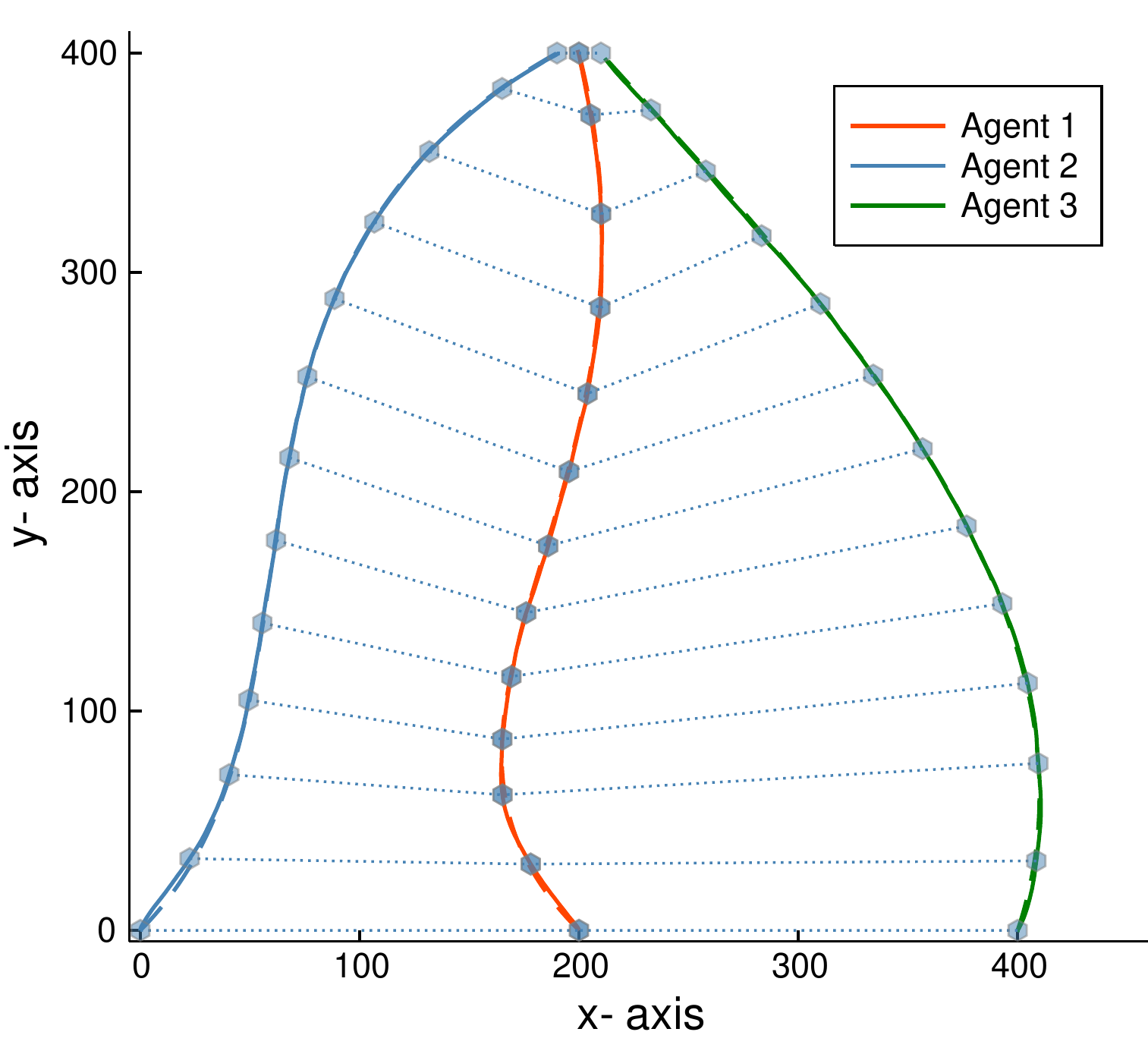}
         \caption{Path following trajectories of three agents are plotted in solid lines. Every three hex points connected by two dotted lines indicate the locations of three agents at the same coordination state.}
         \label{fig:followingwmarker}
     \end{subfigure}
        \caption{Trajectory generation}
        \label{fig:Trajectory generation and time coordination}
\end{figure}

Fig.~\ref{fig:bezier traj generation} shows the trajectories generated by~\eqref{eq:Bezier curve}, and the B\'ezier curve control points for each agent are marked with colored dots.
Agent $i$ aims to follow the trajectory starting from point $\mathbf{P}^{(0)}_{i}$ and plans to arrive at the destination point $\mathbf{P}^{(3)}_{i}$ simultaneously.
To achieve these goals, the time coordination controller proposed in \eqref{eq rb control} is used to update the consensus network in~\eqref{eq consensus network}; then a proportional-derivative (PD) tracking controller is used to track the virtual target generated by the coordination state in~\eqref{eq consensus network}.

The parameters used in~\eqref{eq rb control} and the PD controller were set to the following values:
\begin{align*}
    \rho  = \frac{1}{1200}, \quad k_e = 0.005, \quad k_s = 0.005, \quad k_p = 0.05\quad \text{and} \quad k_i = 0.315,
\end{align*}
where $k_p$ and $k_i$ are the proportional gain and the derivative gain.

Fig.~\ref{fig:followingwmarker} shows the path following and time coordination results.
A series of locations of the three agents are plotted by the hex points.
Their connections by the dotted lines indicates that they have the same coordination states.
We can see that the time coordination and PD control are both well designed, and all of the agents arrived at goal destination simultaneously.

\subsection{In the presence of GPS spoofing attack}
\begin{figure}[ht]
     \centering
     \begin{subfigure}[b]{0.45\textwidth}
         \centering
         \includegraphics[width=\textwidth]{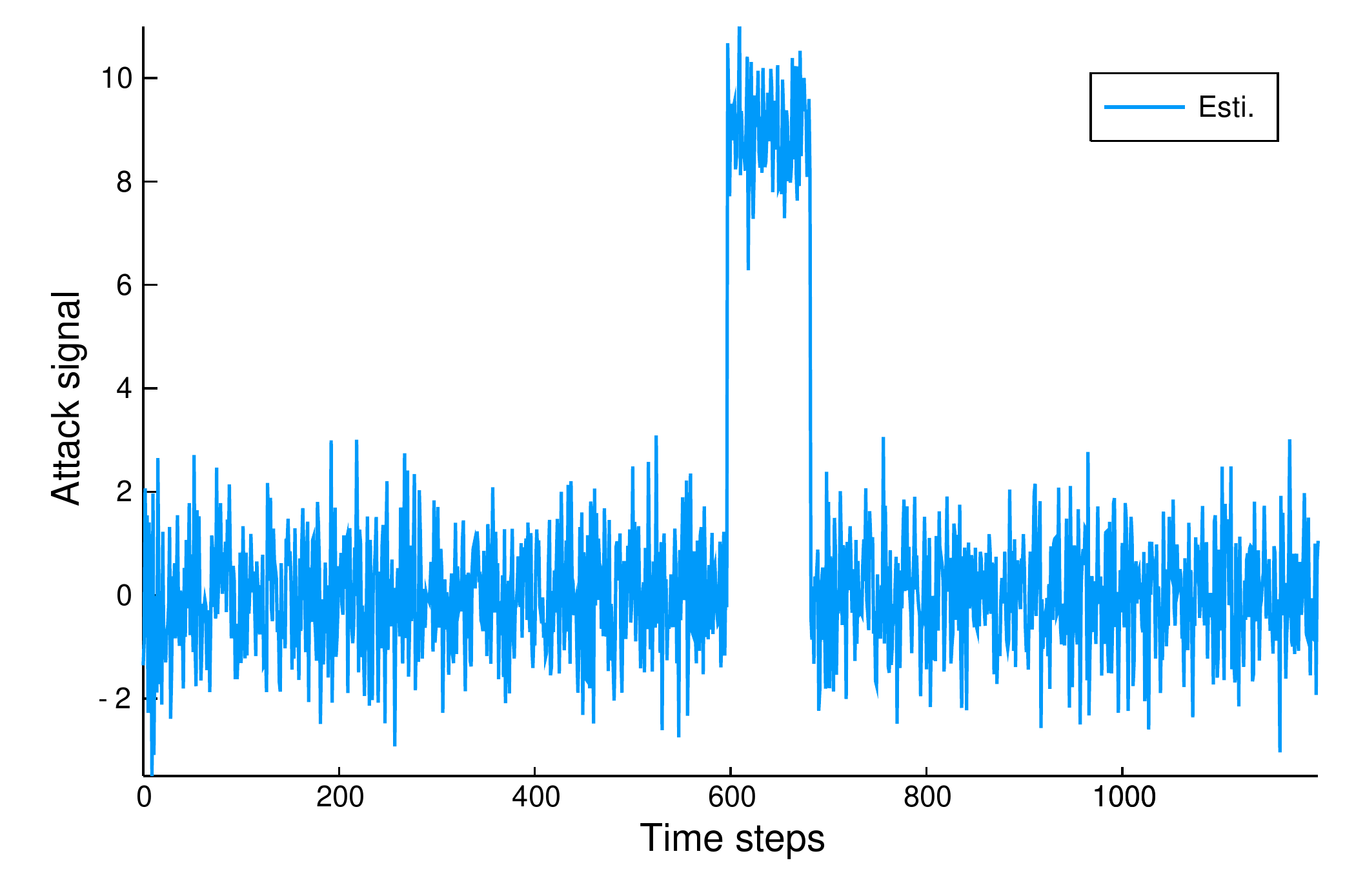}
         \caption{Attack signal estimation.}
         \label{fig:attack esti}
     \end{subfigure}
     \hfill
     \begin{subfigure}[b]{0.45\textwidth}
         \centering
         \includegraphics[width=\textwidth]{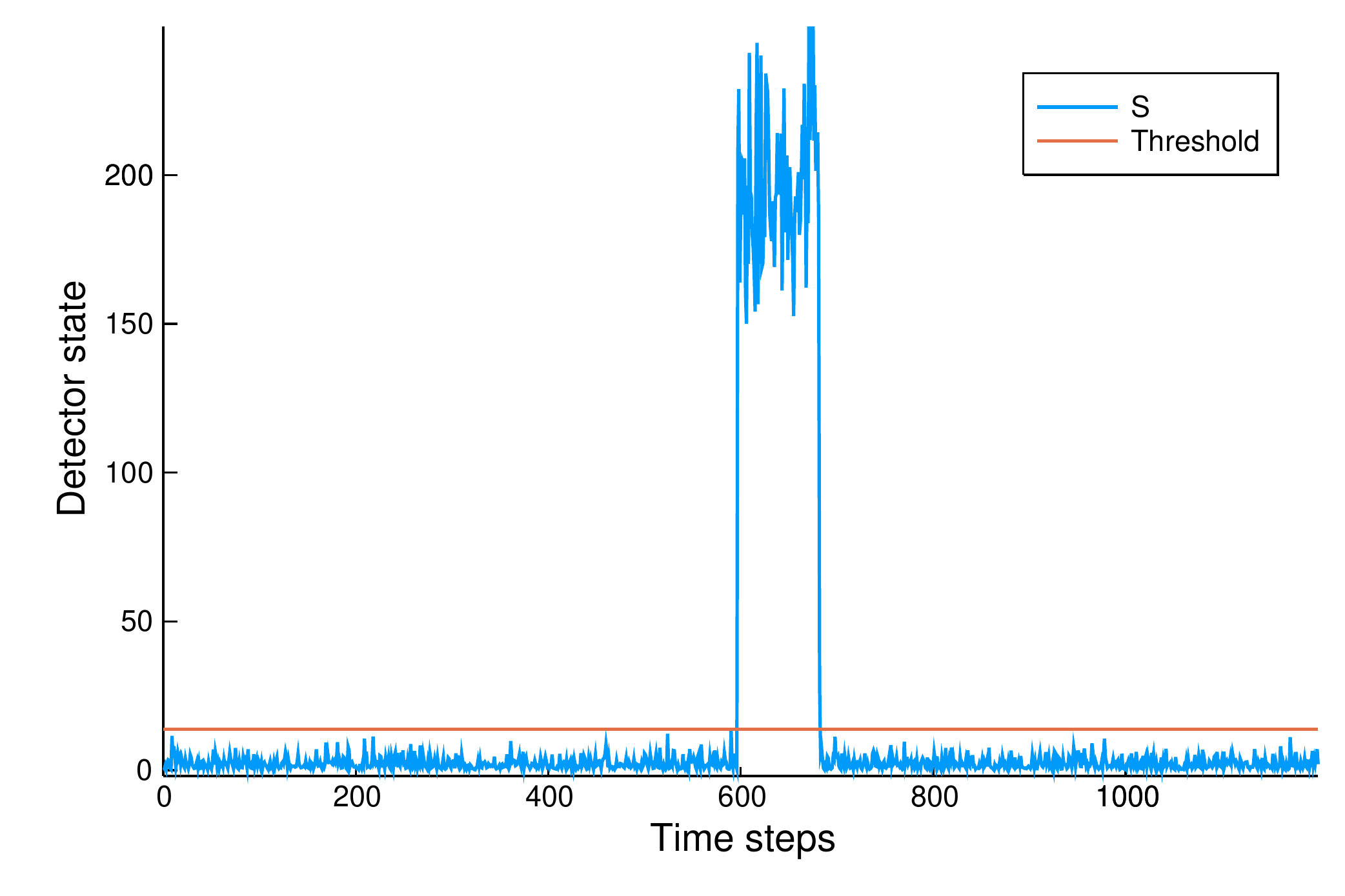}
         \caption{Attack detection. }
         \label{fig: CUSUM}
     \end{subfigure}
        \caption{Attack estimation and detection.}
        \label{fig:three graphs}
\end{figure}

The GPS attack happens when the UAV is in the effective range of the spoofing device.
In this attack scenario, the attack signal is $d_k = [10, 10]^\top$ and the effective range of the spoofing device $r_\textit{effect} =30$.
The location of the attacker is $x^a_k = [200,200]^\top$, which is unknown to the UAV until it is inside the effective range of the spoofing device.
The estimation obtained by~\eqref{est d_k} is shown in Fig.~\ref{fig:attack esti}.  
The detector state $S_k$ can be obtained by using the estimated attack signal as in~\eqref{e003.1}.
The abnormal high detector state values shown in Fig.~\ref{fig: CUSUM} implies that there is an attack.
Statistical significance of the attack is tested using the CUSUM detector described in~\eqref{e003.2} with the significance $\alpha$ at $1\%$.
The threshold is calculated by $\frac{\chi^2_{df}(\alpha)}{1-\delta}$ with $\alpha = 0.01$ and $\delta = 0.15$.

\begin{figure}[ht]
    \centering
    \includegraphics[width=0.4\textwidth]{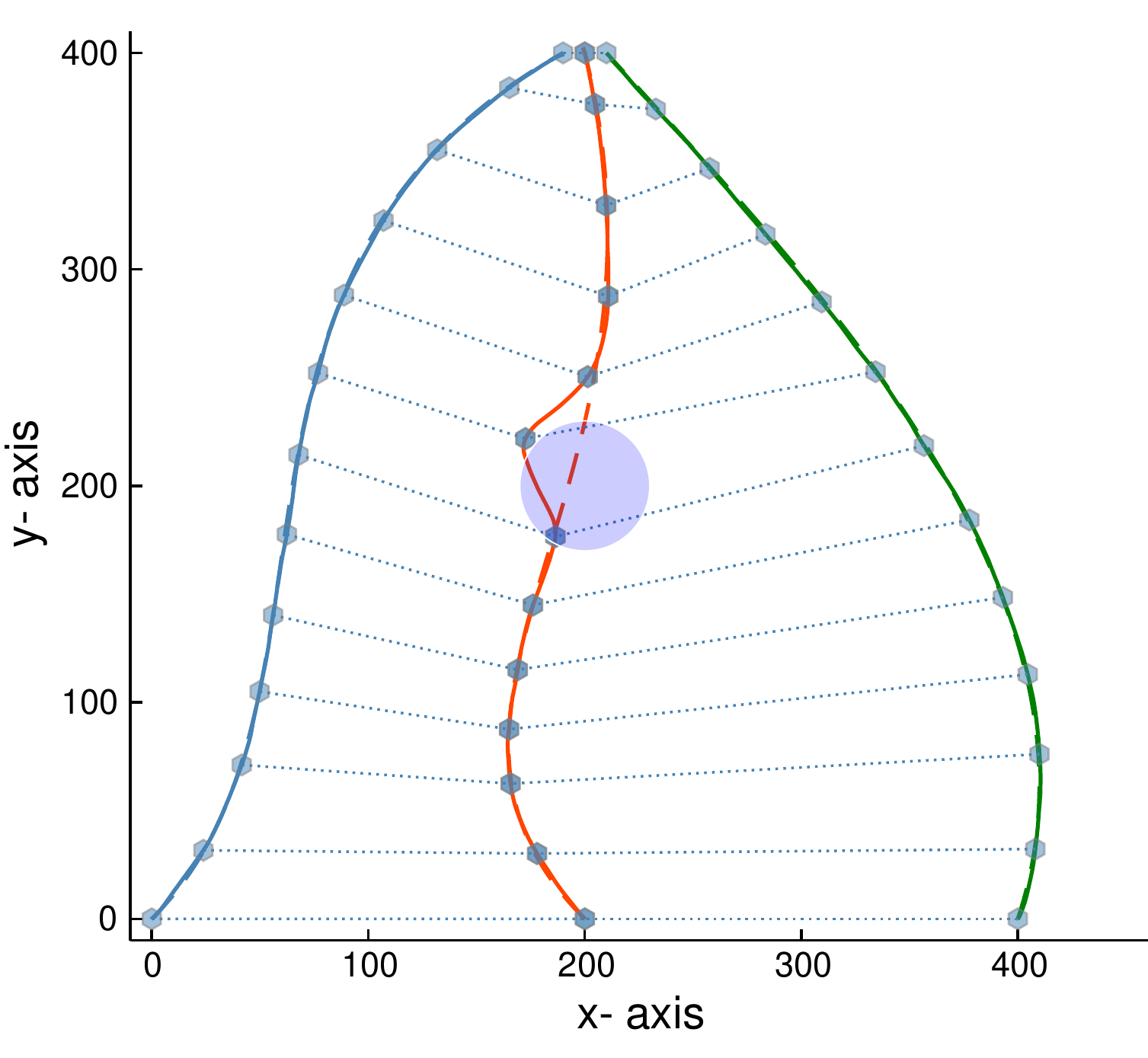}
    \caption{Trajectory in the presence of the attack. The attacker is located at $[200, 200]^\top$ with $r_{\textit{effect}} = 30$, which is displayed as the light blue circle.}
         \label{fig:attack_traj30}
\end{figure}

ESC in Program~\ref{mpc04} with the prediction horizon $N = k^{esc} + 50$ and the scaling parameter $\beta = 10000$ is used to generate the new trajectory for safety operation.
Fig.~\ref{fig:attack_traj30} shows the trajectory of the simulated attack scenario.
ESC drives the attacked UAV away from the effective range of the spoofing device; time coordination is achieved and all of the agents arrive at the destination points simultaneously.

\begin{figure*}[thpb]
        \centering
        \begin{subfigure}[b]{0.22\textwidth}
            \centering
            \includegraphics[width=\textwidth]{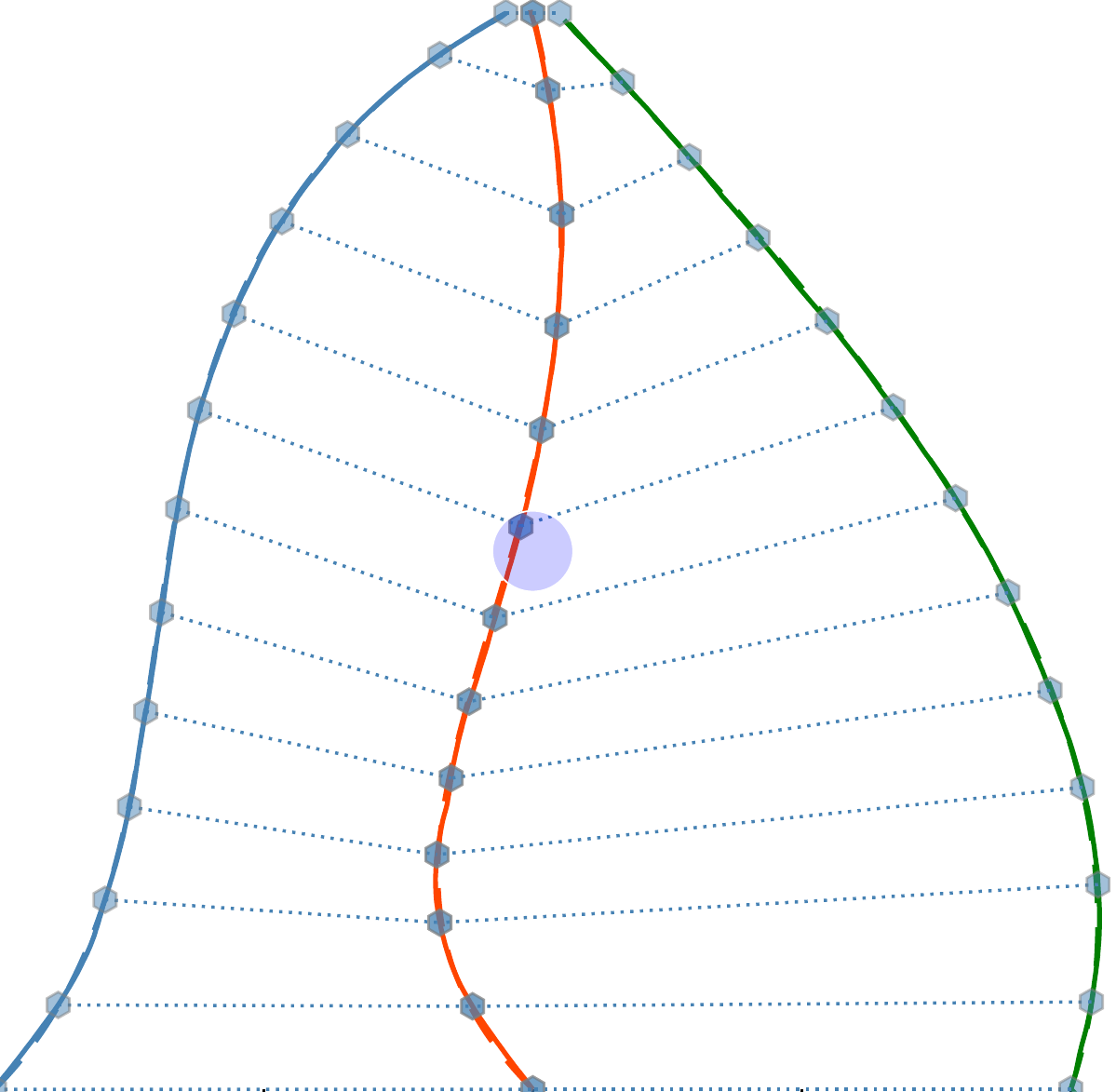}
            \caption[]%
            {{\small $r_{\textit{effect}} = 15$}}    
            \label{fig:mean and std of net14}
        \end{subfigure}
        \hfill
        \begin{subfigure}[b]{0.22\textwidth}  
            \centering 
            \includegraphics[width=\textwidth]{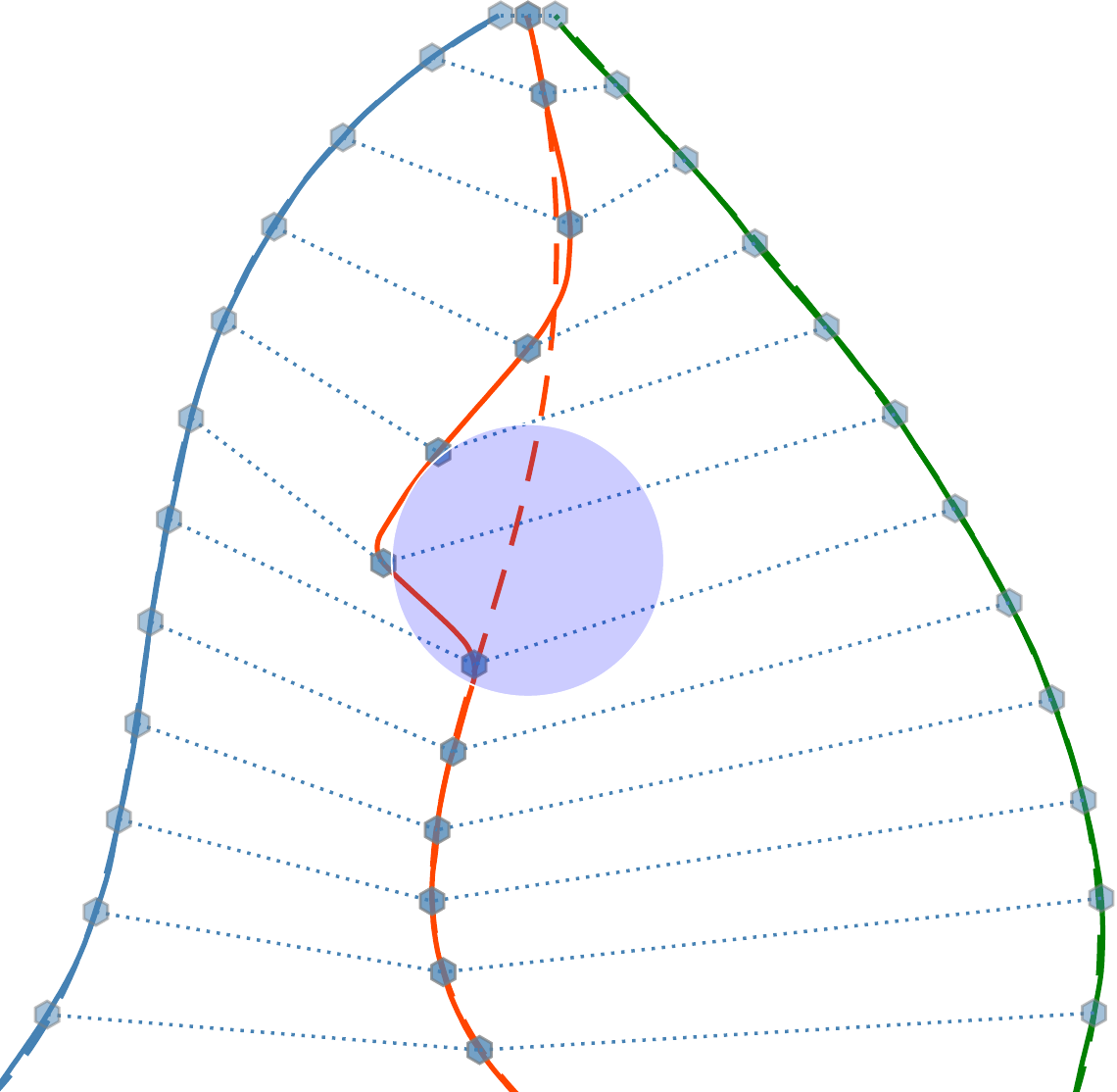}
            \caption[]%
            {{\small $r_{\textit{effect}} = 50$}}    
            \label{fig:mean and std of net24}
        \end{subfigure}
        \hfill
        \begin{subfigure}[b]{0.22\textwidth}   
            \centering 
            \includegraphics[width=\textwidth]{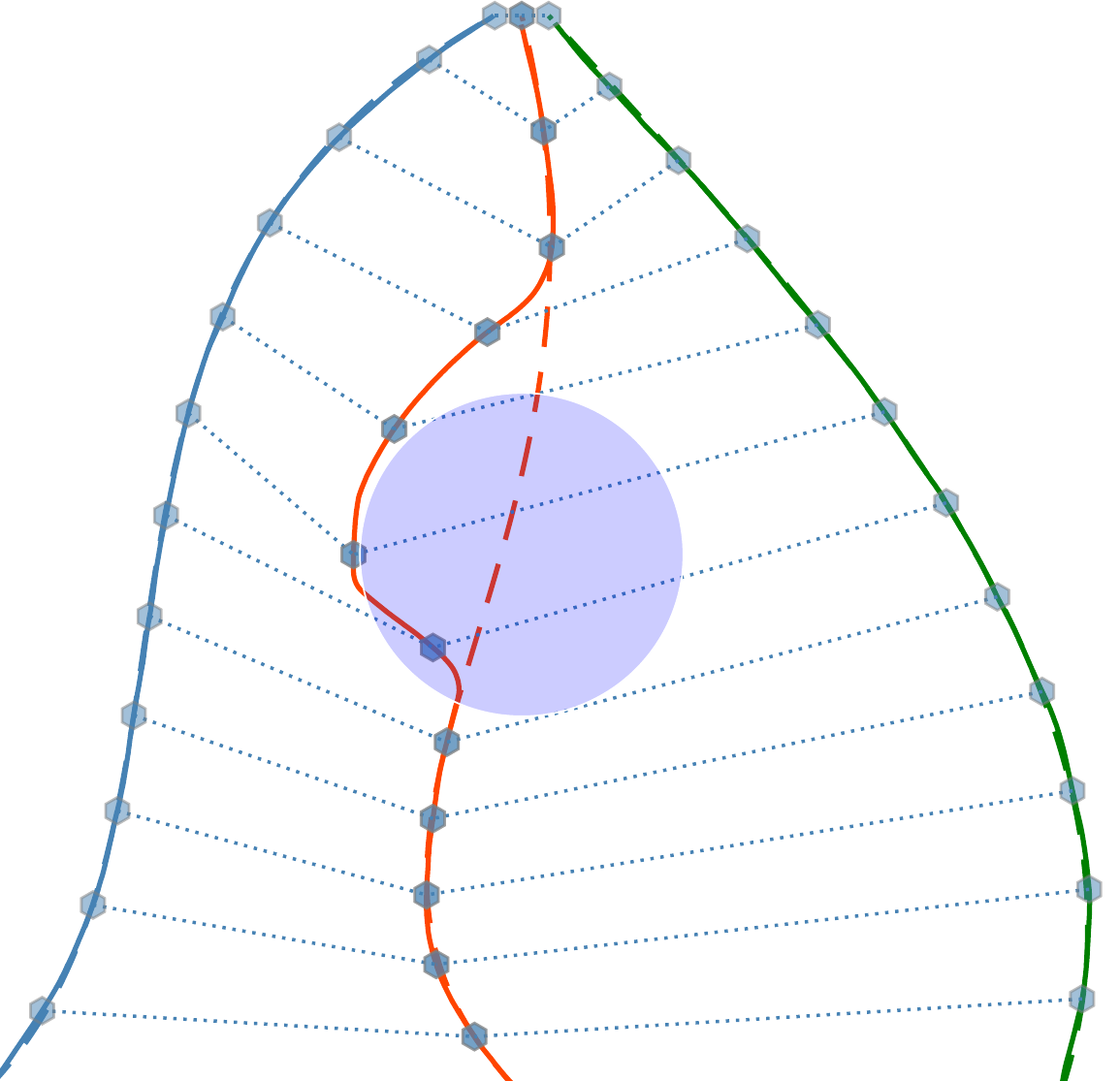}
            \caption[]%
            {{\small $r_{\textit{effect}} = 60$}}    
            \label{fig:mean and std of net34}
        \end{subfigure}
        \quad
        \begin{subfigure}[b]{0.22\textwidth}   
            \centering 
            \includegraphics[width=\textwidth]{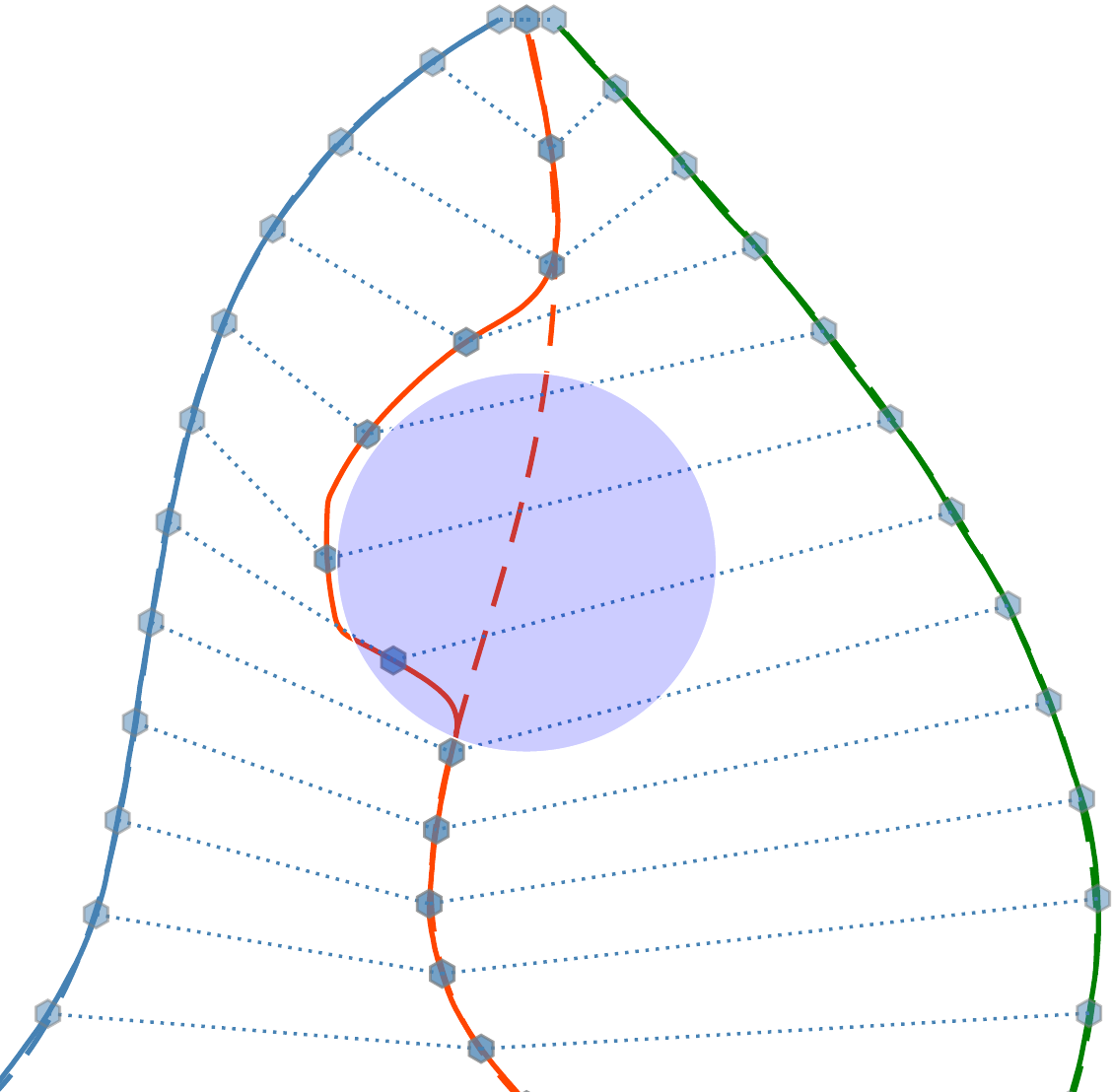}
            \caption[]%
            {{\small $r_{\textit{effect}} = 70$}}    
            \label{fig:mean and std of net44}
        \end{subfigure}
        \caption
        {Trajectories when attacker is located at $[200, 200]^\top$ with different effective ranges.} 
        \label{fig:mean and std of nets}
    \end{figure*}

Fig.~\ref{fig:mean and std of nets} presents how the proposed control framework performs in different cases where $r_{\textit{effect}} \in \{15,50,60,70\}$.
Regardless of the size of $r_{\textit{effect}}$, the UAV will escape the effective range within the escape time and achieve time coordination.
In Fig.~\ref{fig:mean and std of net14}, the attacked UAV can pass the attacker without changing the direction or even its speed, since $r_{\textit{effect}}$ is small enough. From Fig.~\ref{fig:mean and std of net24} to Fig.~\ref{fig:mean and std of net44}, the UAV drives away from the effective range within the escape time and tries to get back to the assigned trajectory.

\section{Conclusion}\label{sec:conclusion}

We presented a safety-constrained bi-level control framework for multi-UAV systems that achieves a consensus of coordination states at the \textit{time-critical coordination level} and adapts the UAV(s) to support resilient state estimation and path re-planning at the \textit{safety-critical control level}.
In particular, the time coordination and the trajectory generation guaranteed the consensus of coordination states and provide the virtual targets for UAV(s) to follow at the first level.
A resilient state estimator has been designed and the $\chi^2$ CUSUM algorithm is used for attack detection.
The state estimation suffers from the increasing variance due to the limited senor availability in the presence of the GPS spoofing attack.
In this case,  the robust controller cannot drive the attacked UAV(s) outside the effective range of the spoofing device with the tolerable estimation errors.
The large estimation errors will cause safety problems and will fail the global path following mission.
To solve these problems, the escape controller (ESC) was designed to escape away from the effective range of the spoofing device within the escape time and complete the global path following mission safely.
The simulations of a three-UAV system were given to demonstrate the results.


\bibliography{z_ref} 
\end{document}